\documentclass[11pt,A4,epsfig]{article}

\usepackage{epsfig}

\def\be{\begin{equation}}
\def\bea{\begin{eqnarray}}
\def\eea{\end{eqnarray}}

\newcommand{\GeV}       {\ensuremath{\mathrm{GeV}} }
\newcommand{\GeVf}       {\ensuremath{\mathrm{GeV}}}

\newcommand{\GeVcc}     {\ensuremath{\mathrm{GeV}/c^2} }

\newcommand{\GeVccf}     {\ensuremath{\mathrm{GeV}/c^2}}

\newcommand{\pbinv}     {\ensuremath{\mathrm{pb}^{-1}} }
\newcommand{\pbinvf}     {\ensuremath{\mathrm{pb}^{-1}}}

\newcommand{\sqrts}     {\ensuremath{\sqrt{s}} }

\newcommand{\ee}{\mbox{$\mathrm{e}^{+}\mathrm{e}^{-}$}}

\newcommand{\hnunu}     {\ensuremath{\mathrm{H}\nu \bar \nu} }
\newcommand{\gam}       {\ensuremath{\gamma \gamma} }
\newcommand{\zg}        {\ensuremath{\mathrm{Z} \gamma} }

\newcommand{\ww}        {\ensuremath{\mathrm{WW}} }
\newcommand{\zz}        {\ensuremath{\mathrm{ZZ}} }

\newcommand{\mz}        {\ensuremath{m_{\mathrm{Z}}} }
\newcommand{\mh}        {\ensuremath{m_{\mathrm{H}}} }
\newcommand{\mhf}        {\ensuremath{m_{\mathrm{H}}}}

\newcommand{\cls}       {\ensuremath{\mathrm{CL}_{s}} }

\begin{document}

\vspace*{-1.8cm}
\begin{flushright}{\bf LAL 00-50}\\
{September 2000}\\
\end{flushright}
\vspace*{1cm}

\begin{center}
{\bf\LARGE STANDARD MODEL HIGGS AT LEP}
\end{center}
\vspace*{0.3cm}
\begin{center}
{\bf\Large Esther Ferrer Ribas}
\end{center}
\begin{center}
{\it\large Laboratoire de l'Acc\'el\'erateur Lin\'eaire}\\
IN2P3-CNRS et Universit\'e de Paris-Sud, BP 34 91898 Orsay cedex, France
\end{center}

\vspace*{0.3cm}

\begin{center}
\parbox{14cm}{
In 1999  the LEP experiments collected data at centre of mass 
energies between 192 and
202~\GeV for about 900~\pbinv integrated luminosity. Combined results
  are presented for the search for the Standard Model Higgs boson.
No statistically significant excess has  been observed when compared
to Standard Model background expectation which can be translated into a lower
bound on the mass of the Higgs boson at 107.9~\GeVcc at 95~\%
confidence level.}
\end{center}
\vspace*{0.3cm}

We present combined results from the ALEPH, DELPHI, L3 and OPAL 
Collaborations on the search for  the
Standard Model Higgs boson. The  results are obtained by  combining 
the results  from  data
collected in  1999 at centre-of-mass energies between 192 and 
202~\GeV  and earlier  data collected at lower
energies~\cite{direct189}.

\section{The Higgs boson in the Standard Model }
The experimental observation of the Higgs boson(s) would be of great importance
for the understanding of the spontaneous breaking of the electroweak
symmetry.  In the Standard Model, one scalar neutral boson is 
introduced. The mass of this boson,
  \mh, is a free parameter of the theory.

\begin{figure}[h]
\begin{center}
\epsfig{figure=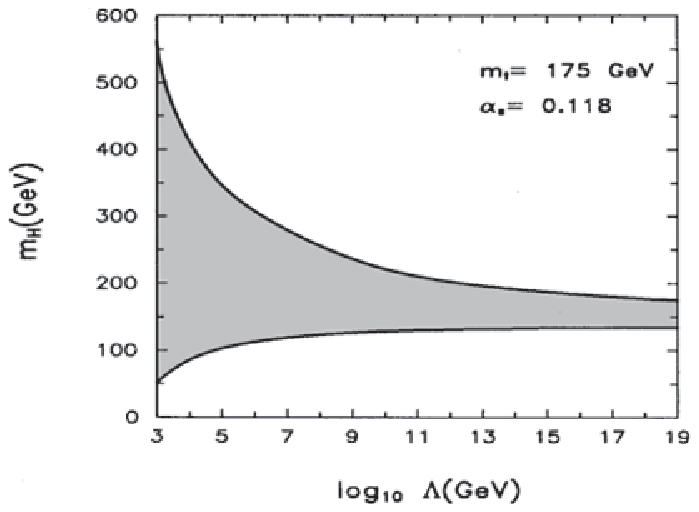}
\epsfig{figure=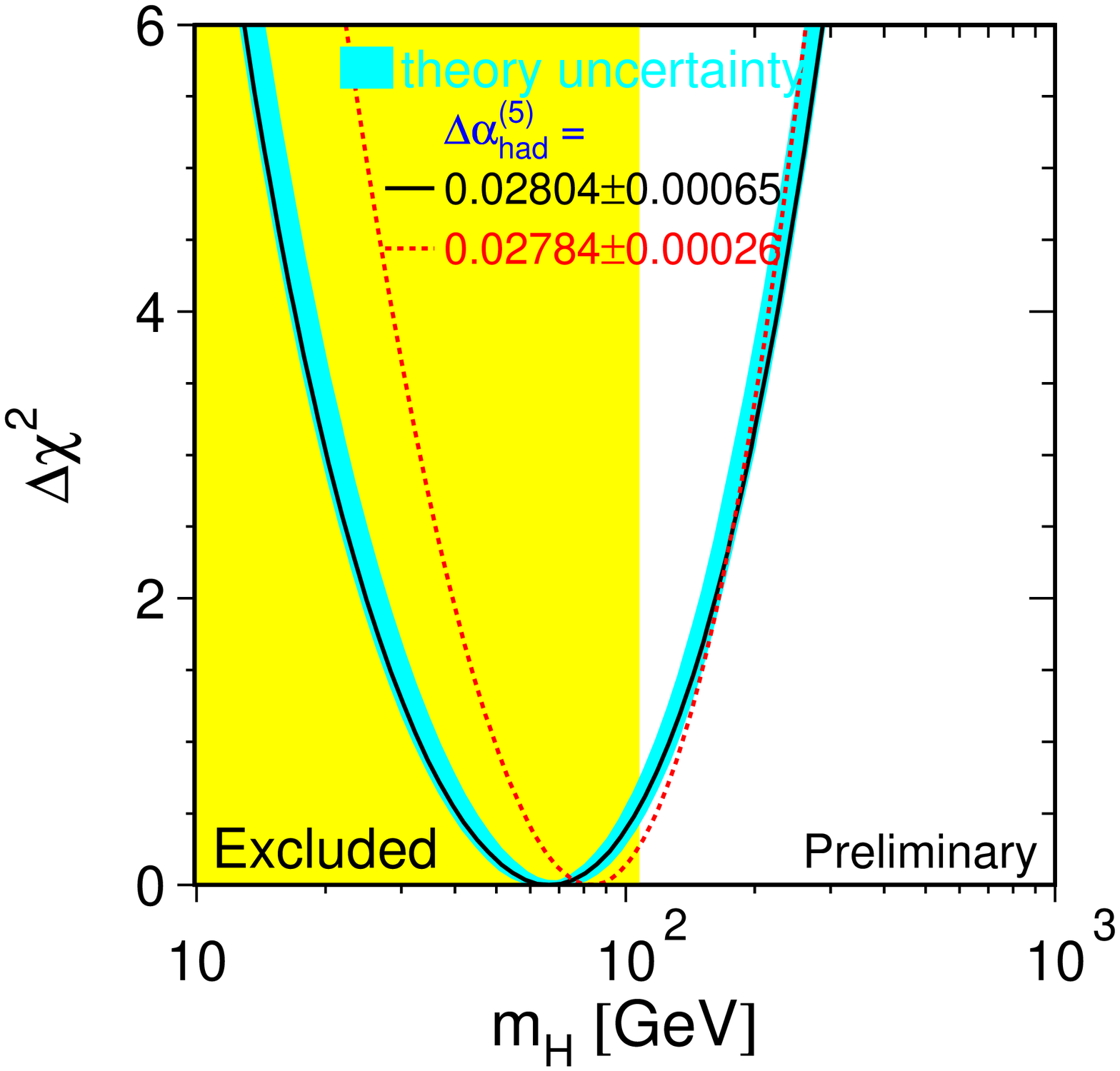,width=8cm}\\
(a)\hspace{6cm}(b)\\
\caption{(a) Constraints on the Higgs mass by stability of the 
vacuum and triviality arguments
as a function of the $\Lambda$ scale for  a given of the top 
mass ($m_{\mathrm{t}}=175~\GeVcc$). The grey region between the curves
shows the allowed zone by triviality and vacuum stability arguments.
(b) $\chi^{2}$ variation as a function of the Higgs mass from the 
global fit of the electroweak measurements.}
\end{center}
\label{fig:mhconstraints}
\end{figure}

However by theoretical arguments, constraints on \mh can be  derived. On one
hand, triviality arguments give rise to an upper bound on \mh as a function
of the  scale $\Lambda$ and the top mass. On the  other hand, vacuum 
stability will
give a lower bound as a function of the top mass. In 
Figure~\ref{fig:mhconstraints}(a) the bounds
on \mh given by the triviality and vacuum stability arguments are 
shown as a function
of $\Lambda$ at a given top mass~\cite{achille}.

The electroweak observables are measured with a high precision (of
the order of $^{0}/_{000}$). This precision allows us to be sensitive to 
the radiative corrections due to the top quark and to the Higgs boson
via the loop diagrams. In order to minimise the uncertainties on
the theoretical values, the best measured observables are used~: G$_{F}$,
$\alpha_{\mathrm{QED}}$ and \mz. For given values of G$_{F}$,
 $\alpha_{\mathrm{QED}}$ and \mz indirect measurements for $m_\mathrm{t}$ 
and \mh can be extracted. The precision will be better on $m_\mathrm{t}$ than
on $m_{\mathrm{H}}$, thanks to the quadratic dependence 
($\sim m_{\mathrm{t}}^2$) whereas
for \mh the dependence is logarithmic ($\sim \ln\mh^2$). All of
the electroweak measurements are adjusted in a global fit. The result on
$m_\mathrm{t}$ is one of the biggest successes of the Standard Model giving
an indirect measurement ($m_{\mathrm{t}} = 174.2 ^{+10.9}_{-6.9}~\GeVcc$)
in agreement with the direct measurement 
($m_{\mathrm{t}}= 174.3 \pm 5.1~\GeVcc$)
 and only half its precision. In the same way, the mass of the Higgs boson 
is expected~\cite{elwmes} to be
smaller than 188~\GeVcc at 95~\% CL as can be seen from 
Figure~\ref{fig:mhconstraints}b.

Direct searches obtained a 95\% CL  lower  bound of 95.2~\GeVcc from 
the combination of
  earlier data  collected by the LEP experiments  at centre-of-mass 
energies up to 189~\GeVf~\cite{direct189}.
\newpage

\section{Production and topologies}
At LEP2 energies the dominant Higgs boson production is the Higgsstrahlung
process $\ee\rightarrow$ HZ where a Higgs boson is produced in 
association with a
Z boson\footnote{The fusion process in which a Higgs boson is 
produced by the fusion
of two W bosons has considerable smaller cross section at LEP energies. It 
becomes competitive
at the kinematical limit ($\mh \ge \sqrts-\mz$). This process contributes 
only for the H$\nu \bar \nu$
final states and gives the possibility of exploring mass regions beyond the
kinematical limit.}.

  The decays of the Z and the Higgs boson determine the final topology.
For Higgs masses below 120~\GeVccf, the Higgs boson decays predominantly into a pair of b quarks ($\sim85\%$) followed  by the decay mode into $\tau^+\tau-$ ($\sim8\%$). The remaining decays concern pairs of c quarks or a pair of gluons.
The Z boson is a much more democratic particle and decays 70\% into a 
quark-antiquark pair 
 ( shared in about 20\% for d-type quarks and 15\% for u-type quark),
20\% into a pair of neutrino-antineutrino and the rest in a pair of leptons.

The search for the Higgs boson consists of the following final 
states: H$\mathrm{q}\bar\mathrm{q}$
  ($\sim 60\%$), H$\nu \bar \nu$ ($\sim17\%$), H$l^+l^-$ ($\sim9\%$) 
and the rest mainly
$\tau^+\tau^-\mathrm{q} \bar \mathrm{q} $ and $\mathrm{q} \bar{ \mathrm{q}} 
\tau^+ \tau^-$.  All in all about 93\% of the Higgs
 decay channels are studied.

\begin{figure}[h]
\vspace*{-0.4cm}
\begin{center}
\epsfig{figure=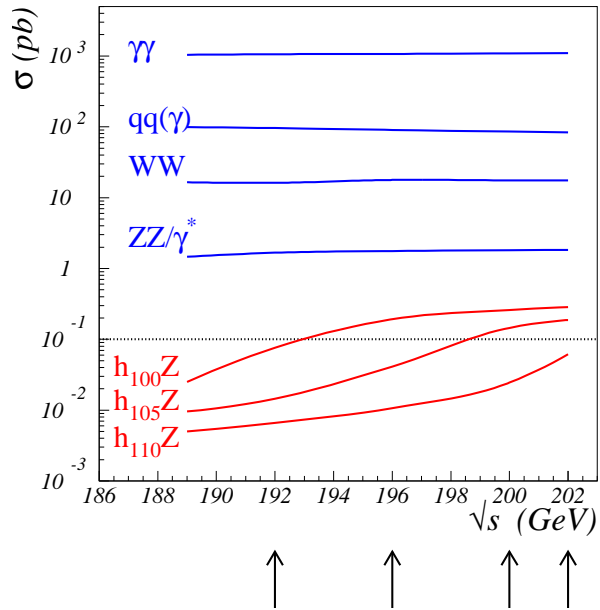,width=8cm}
\vspace*{0.5cm}
\caption[]{
Evolution of the cross sections of the main background process as a function of
the centre of mass energies. The arrows show the centre of mass 
energies where data
has been  collected during 1999. The cross sections for 
Higgsstrahlung production
for \mh=100, 105 and 110~\GeVcc are also shown. The dotted line shows 
where 10 Higgs
boson would be produced with 100~\pbinv.}
\end{center}
\label{fig:bkg}
\end{figure}

The background processes can have cross sections of order of
magnitude greater than those for the Higgs production.
The cross sections of the main background processes as a function of \sqrts
  are shown in Figure~\ref{fig:bkg} as well as the cross sections of 
the HZ signal for different
$m_{\mathrm{H}}$. The \gam events can be easily reduced as their topology is very 
different from the
signal. The \zg and \ww events will be the main backgrounds for all 
channels. The
\zz events with final states containing a pair of b quarks are an 
irreducible background
at  $\sqrts=189~\GeV$. At centre-of-mass energies greater  than
192~\GeVf, the sensitivity of  the  Higgs searches being beyond 
$m_{\mathrm{Z}}$, there is some discrimination in mass 
between events coming from 
HZ or \zz production. It is clear from this figure, when comparing 
background and signal cross sections, that the event selections have
to reduce background event contamination by a few orders of magnitude.

\vspace*{-0.6cm}
\section{Analysis strategies}

The analyses used by the four LEP collaborations are described in 
individual documents~\cite{neutral-aleph,neutral-delphi,neutral-l3,neutral-opal}. We give 
a brief overview
of the common features.

The event selections rely heavily on the b-tagging algorithm  as 85\% of Higgs
boson decays are $\mathrm{b}\bar\mathrm{b}$. Typical performances of
these algorithms are a 90\% b-jet efficiency selection
for  a light  quark reduction of about 80\%.

The analyses are often divided into two main steps: preselection cuts 
and discriminant variable.
In  the preselection steps, the aim is to reduce events coming from 
background processes with
  different topological characteristics from signal events. The \gam 
background can be
reduced to a negligible level and the 2-fermion and 4-fermion events 
are reduced by about
a factor 15, keeping a high efficiency for the signal events.

\begin{figure}
\begin{center}
ALEPH\\
\epsfig{figure=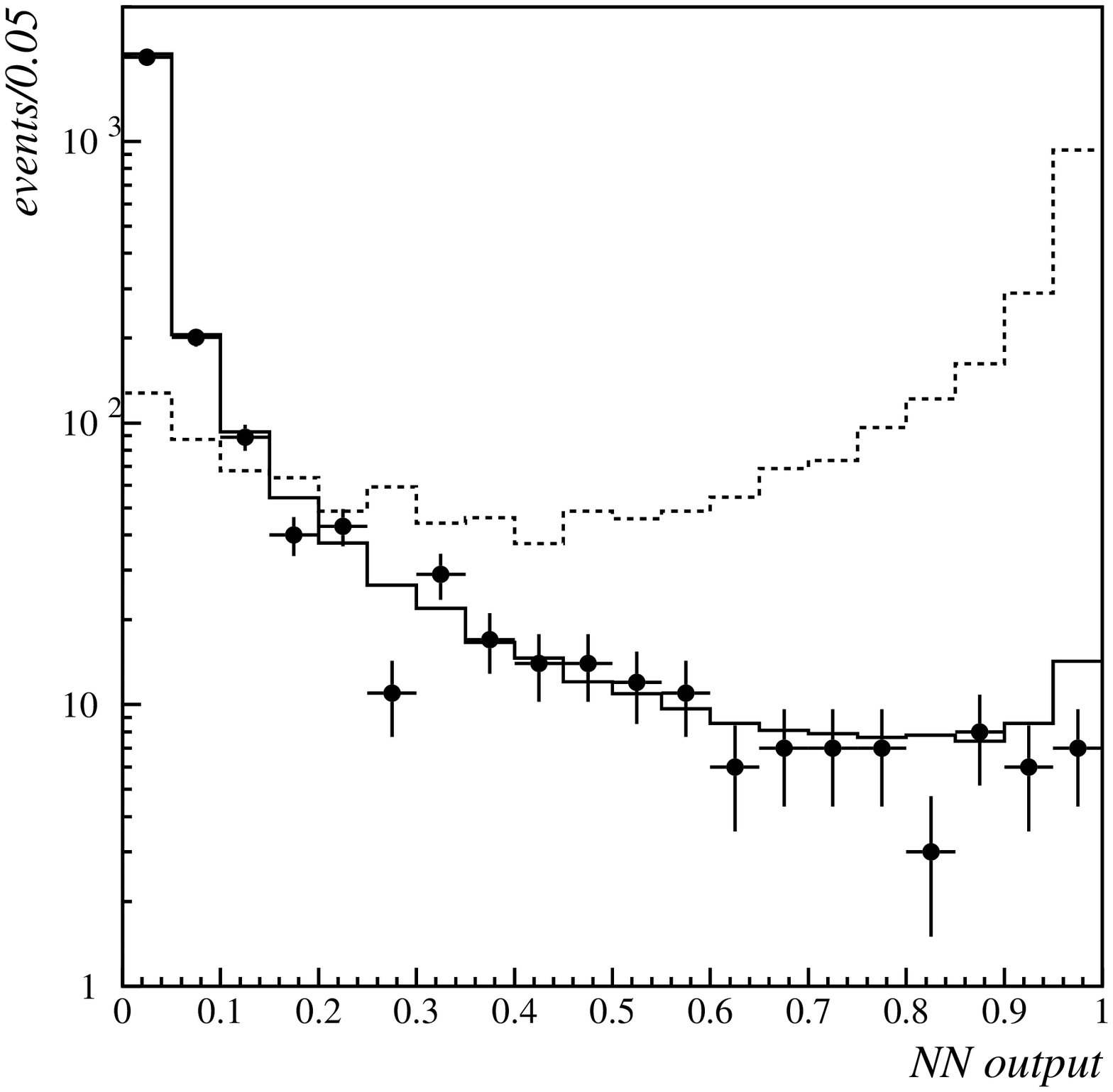,width=0.5\textwidth}\\
(a)\\
\vspace{0.7cm}
DELPHI\\
\epsfig{figure=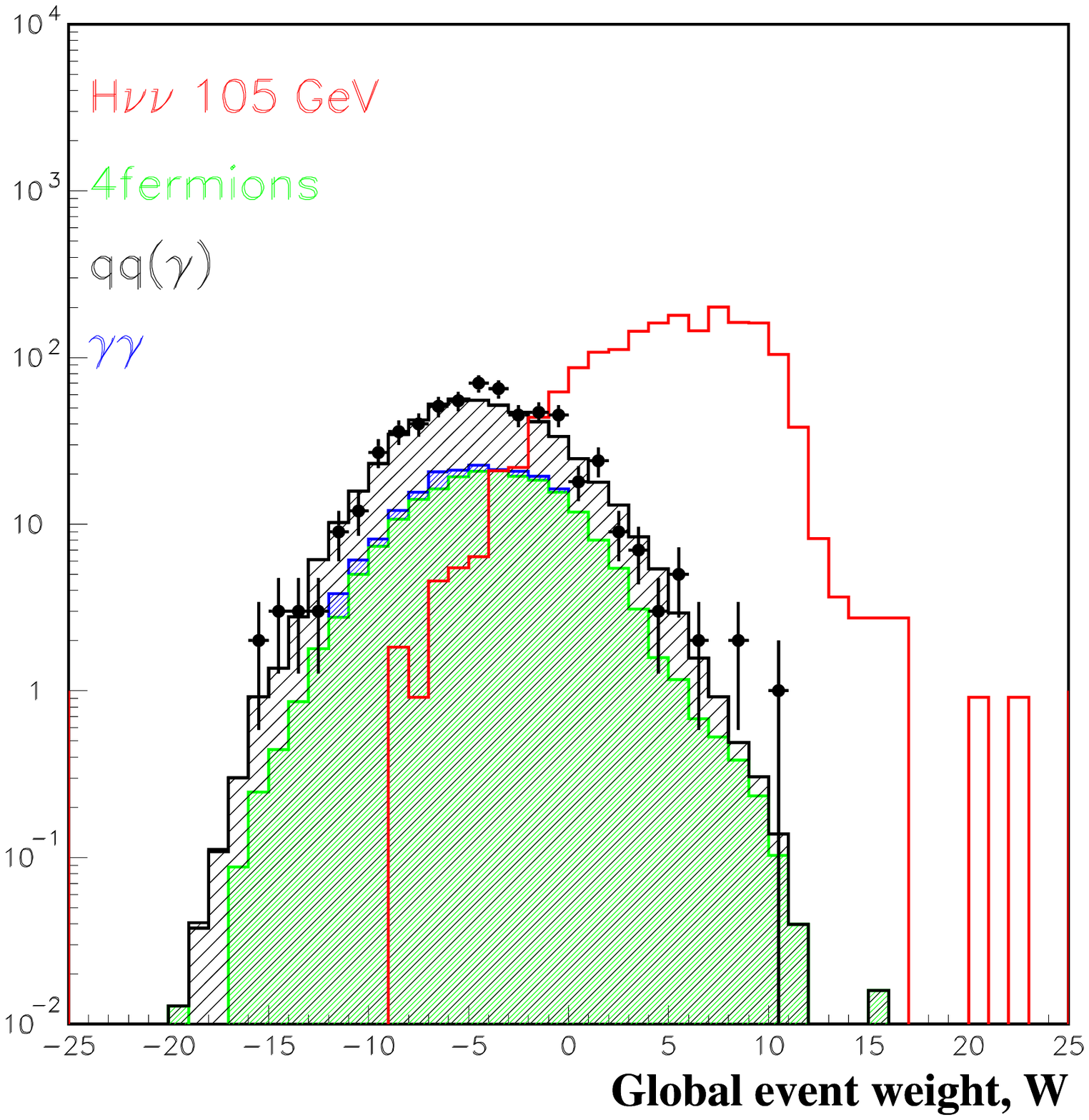,width=0.45\textwidth}
\epsfig{figure=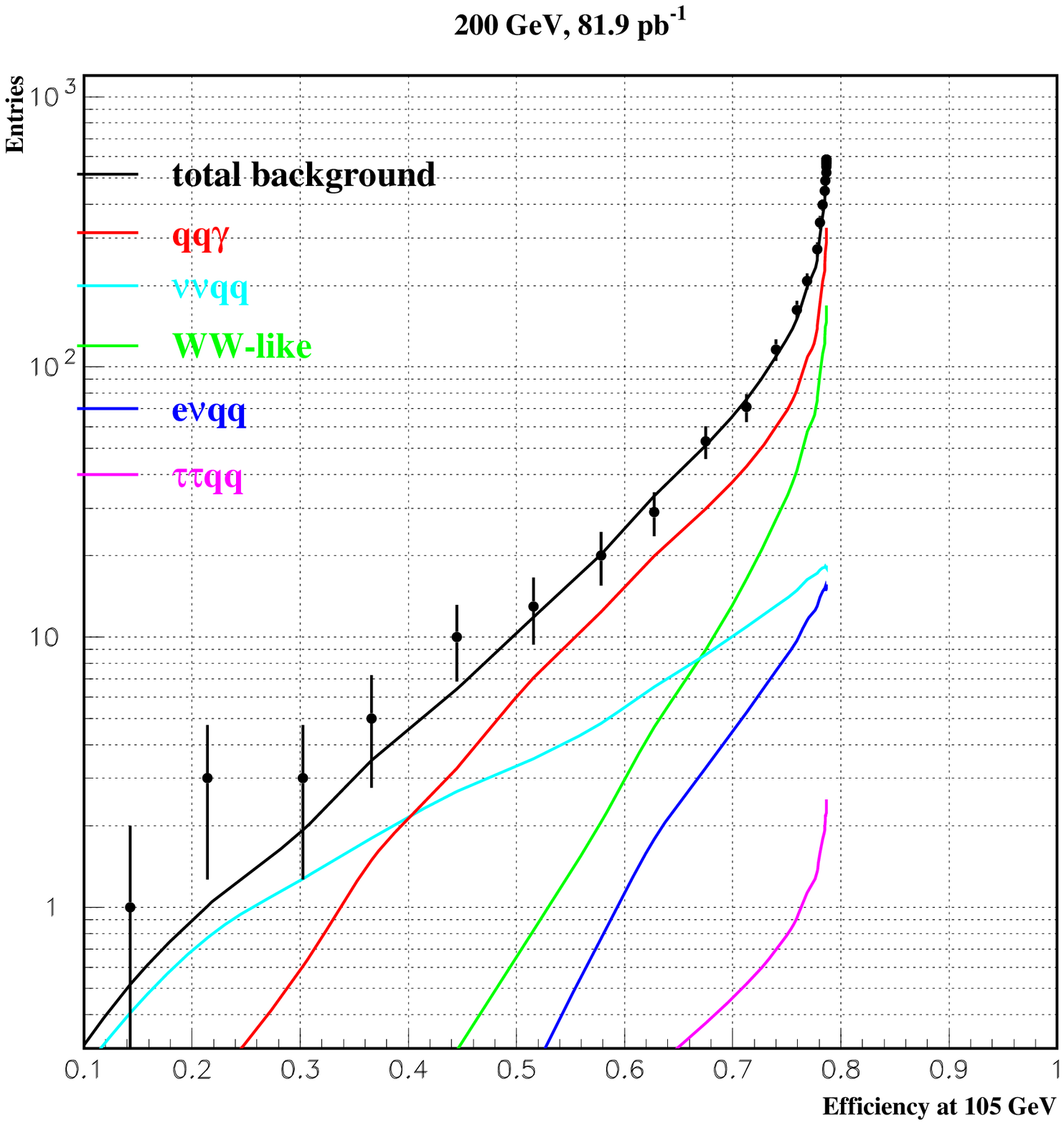,width=0.45\textwidth}\\
(b)\hspace{7cm}(c)\\
\vspace*{0.8cm}
\caption[]{
On the top, (a), distribution of the output of the neural
 network for the four jets
channel for the background contribution (empty histogram), signal
(dotted unnormalised histogram) and real data (points) for all the
data recorded during 1999. On the bottom left, (b), distribution of 
the discriminant variable in the missing energy channel for $\sqrts=200~\GeV$
and on the bottom right, (c), distribution of the evolution
of the number of background events as a function of signal efficiency
($\mh=105~\GeVcc$) for the same analysis.}
\label{fig:discri}
\end{center}
\end{figure}

The following step consists of constructing a single variable that 
combines the most
discriminating observables between background and signal events by means of a
likelihood ratio or neural network techniques. One example of this kind
of variable is shown in Figure~\ref{fig:discri}(a) where the output 
of the ALEPH 
neural network for the
four jets channels using all 99 data is shown for signal, background and
real data. The separation of the signal and the background events is
remarkable and signal events can be selected by cutting at high
values of the discriminant variable. Figure~\ref{fig:discri}(b)
 shows the discriminant  variable for the
$\hnunu$ analysis in DELPHI. When cutting the discriminant variable
from left to right, the curve~\ref{fig:discri}(c) is obtained 
representing the evolution of the number of events as a function
of the Higgs signal efficiency.

It is noticeable that the background contributions have been reduced
to have a signal over background ratio of the order of unity for an
efficiency of about 30\%.

\section{Results}
\begin{figure}[h]
\begin{center}
\epsfig{figure=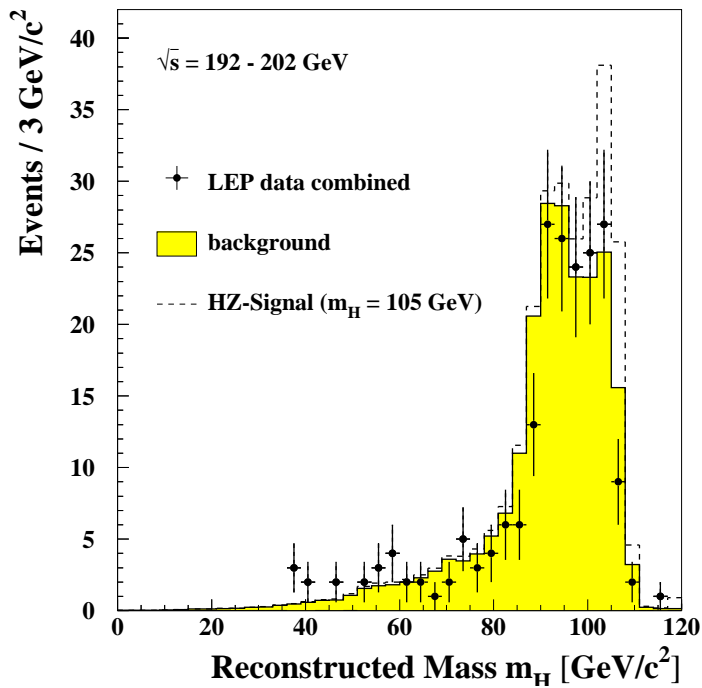,width=10cm}
\caption[]{
LEP-combined distribution of the reconstructed SM Higgs boson mass in
searches conducted at \sqrts between 192 and 202 GeV.  The figure
displays the data (dots with error bars), the predicted SM background
(shaded histogram) and the prediction for a Higgs boson of 105~\GeVcc
mass (dashed histogram).  The figure has been obtained with the
supplementary requirement that the contributions from the four
experiments (selecting the most signal-like set of events) be roughly
equal.  The number of data events selected for this figure is 201
while 220 are expected from SM background processes. A signal at
a Higgs mass of 105~\GeVcc would contribute with 40.7 events.}
\label{fig:sm-mass}
\end{center}
\end{figure}

All data have been analysed by the four experiments. A summary is 
given in Table
~\ref{table-hZ-input}. No statistically significant excess has been observed.
Therefore these results can be translated into an exclusion limit on 
the mass of the Higgs boson.
The observed and expected  limits per  experiment are given in Table~\ref{table-hZ-input}. These
results are combined. As an illustration the distribution
of the reconstructed  mass of the Higgs boson for all channels and 
all experiments
is shown in Figure~\ref{fig:sm-mass}. The figure has been obtained by selecting 
the most similar set
of events and requiring that the contributions from the four experiments be
roughly equal. The number of selected data events is 201 while 220 
are expected  from SM
background processes. In the presence of a signal at 105~\GeVcc mass, 
40.7 events  would
be expected.

\begin{table}[p]
\begin{center}
\begin{tabular}{||l||c|c|c|c||}
\hline\hline
Experiment:                     & ALEPH & DELPHI & L3  & OPAL \\
\hline\hline
192 GeV: Integrated luminosity (\pbinv): &28.9 &25.2-25.9  &29.7 
&28.7-28.9   \\
\phantom{.....}Backg. predicted / Evts. observed 
&&&&  \\
\phantom{..........}Four-jet:              &4.8/3 &19.2/16 &0.7/1 &3.6/5    \\
\phantom{..........}Missing-energy:        &1.5/1 &12.1/13 &0.2/0 &1.5/0     \\
\phantom{..........}Leptonic (e, $\mu$):   &2.8/3 &2.5/1   &0.0/0 &1.0/1    \\
\phantom{..........}Tau channels:          &1.2/1 &0.8/0   &0.0/0 &0.8/1     \\
\hline
196 GeV: Integrated luminosity (\pbinv): &79.9 &74.8-76.9 &83.7 &73.9-74.8  \\
\phantom{.....}Backg. predicted / Evts. observed 
&&&&  \\
\phantom{..........}Four-jet:              &14.8/8 &59.3/51 &5.4/8 
&10.0/17    \\
\phantom{..........}Missing-energy:        &3.8/4 &32.8/32  &1.3/0 
&3.2/2     \\
\phantom{..........}Leptonic (e, $\mu$):   &8.9/4 &6.8/7    &0.2/0 &2.4/2    \\
\phantom{..........}Tau channels:          &3.7/1 &2.4/3    &0.1/0 
&2.1/0     \\
\hline
200 GeV: Integrated luminosity (\pbinv): &86.3 &81.9-84.3 &82.8 &74.8-77.2  \\
\phantom{.....}Backg. predicted / Evts. observed 
&&&&  \\
\phantom{..........}Four-jet:              &17.5/16 &67.2/61 &22.6/24 
&9.3/9    \\
\phantom{..........}Missing-energy:        &3.8/1 &36.6/32   &4.3/7 
&2.8/2     \\
\phantom{..........}Leptonic (e, $\mu$):   &11.2/13 &8.1/9   &1.1/1 
&3.3/5    \\
\phantom{..........}Tau channels:          &4.7/7 &2.6/3     &0.8/0 
&2.7/3     \\
\hline
202 GeV: Integrated luminosity (\pbinv): &41.9 &40.0-41.1 &37.0 &35.2-36.1  \\
\phantom{.....}Backg. predicted / Evts. observed 
&&&&  \\
\phantom{..........}Four-jet:              &9.3/3 &33.8/33 &9.4/14 &4.8/2    \\
\phantom{..........}Missing-energy:        &1.8/1 &17.9/20 &2.8/4 
&1.8/2     \\
\phantom{..........}Leptonic (e, $\mu$):   &5.6/6 &4.2/1   &0.5/0 &1.4/2    \\
\phantom{..........}Tau channels:          &2.4/2 &1.2/0   &0.4/0 &0.8/0     \\
\hline\hline
Total:~~~Integrated luminosity (\pbinv): &237.0 &222-228 &232.4 &213-217  \\
\phantom{.....}Backg. predicted / Evts. observed 
&&&&  \\
\phantom{..........}Four-jet:              &46.4/30 &179.5/161 
&38.1/47 &27.7/33    \\
\phantom{..........}Missing-energy:        &11.0/7 &99.4/97 
&8.6/11 &9.3/6     \\
\phantom{..........}Leptonic (e, $\mu$):   &28.5/26 &21.6/18   &1.8/1 
&8.1/10    \\
\phantom{..........}Tau channels:          &11.9/11 &7.0/6     &1.3/0 
&6.4/4     \\
\hline\hline
Events in all channels &97.8/74 &307.5/282 &49.8/59 &51.5/53   \\
\hline
Limit (\GeVcc) exp. (median) at 95\% CL: &107.7(*) &106.3 &105.3 &105.2 \\
Limit (\GeVcc) observed at 95\% CL: &107.7 &103.9 &106.0 &103.0 \\
\hline\hline
\end{tabular}
\end{center}
\vspace*{1cm}
\caption{ Information related to the searches of the four LEP
experiments for the SM
Higgs boson at energies between 192 and 202~GeV.
In the L3 analysis
the event selection, and thus the expected background and observed
number of events, depend on the Higgs boson mass hypothesis; they are
given here for \mh=105~\GeVccf. (*) In the ALEPH publication the expected
mean is quoted (which is 106.8~\GeVccf) rather than the
median. Also, the confidence level estimator $CL_s$ used by ALEPH
is different from the one used by the other collaborations, shifting
the expected limit of ALEPH upwards by about 0.5~\GeVccf.}
\label{table-hZ-input}
\end{table}

\begin{figure}[h]
\vspace*{0.5cm}
\begin{center}
\epsfig{figure=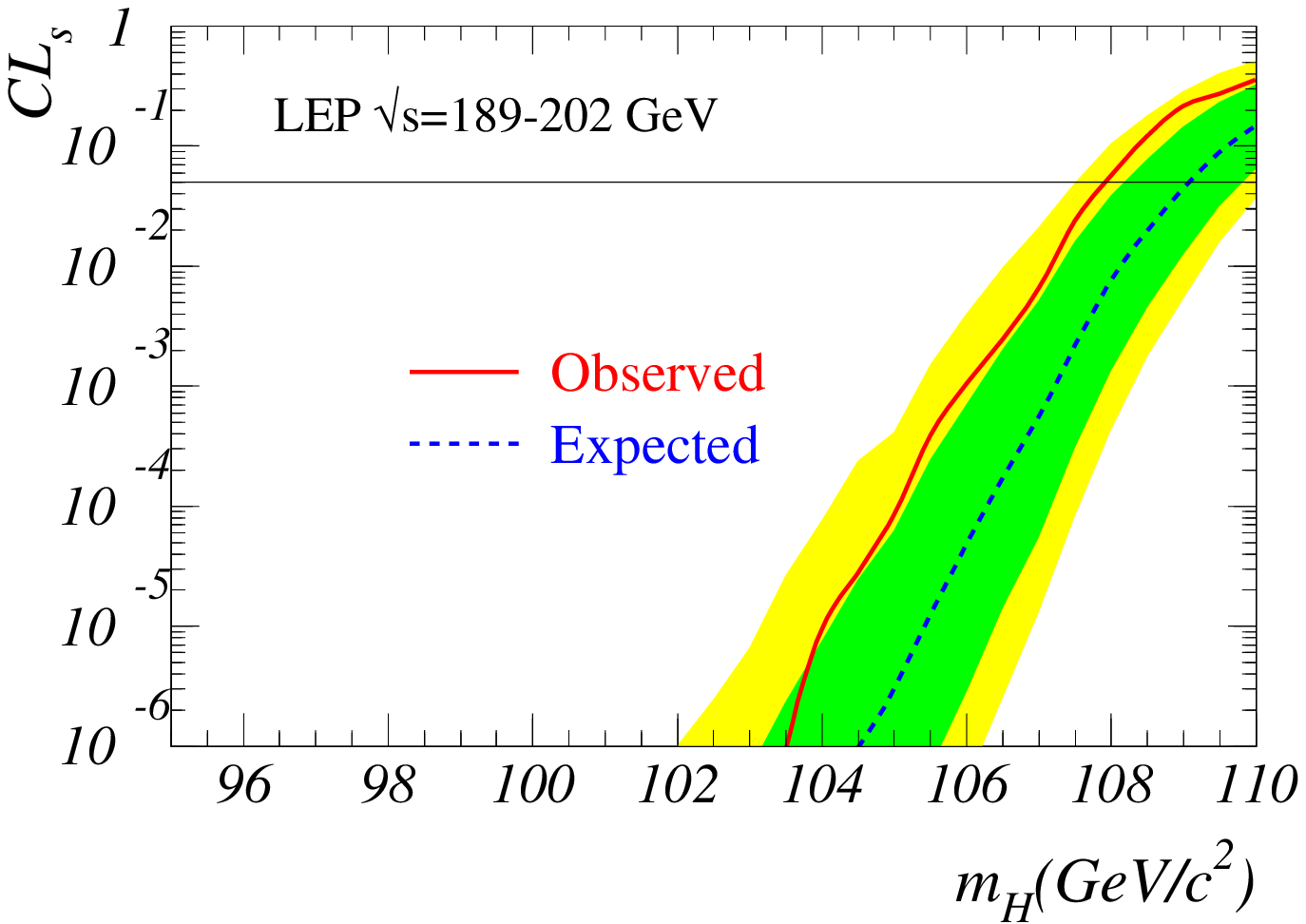,width=0.49\textwidth}
\epsfig{figure=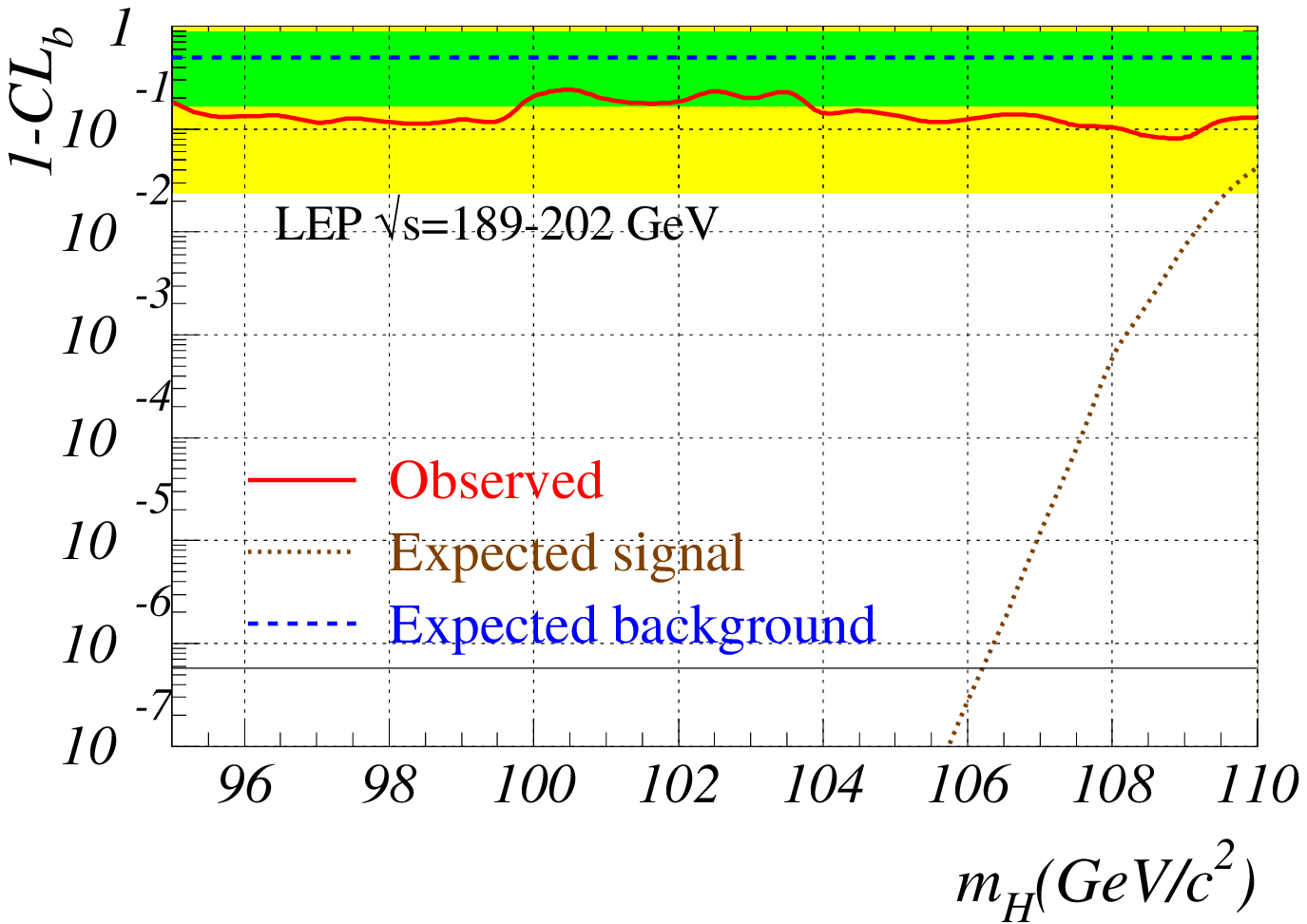,width=0.49\textwidth}\\
\caption[]{
The confidence levels $Cl_s$ (left) and $1-Cl_b$ (right) as a function 
of \mhf. The
  solid curve is the observed result, the dashed curve the median 
result expected
in the absence of a signal. The shaded areas represent the symmetric 
$1\sigma$ and
$2\sigma$ probability bands of $CL_s$ in the absence of a signal. The 
intersections
of the curves with the horizontal line at $CL_s=0.05$ give the mass limits
at the 95\% confidence level. On the left plot the dotted curve shows the
median result expected for a signal of mass given in abscissa. The
horizontal line at $5.7\times 10^{-7}$ indicates the level for a 5$\sigma$
 discovery.}
\label{sm-cls}
\end{center}
\end{figure}

A combined limit can be extracted. The compatibility with background 
of the result is given by $1-CL_b$,
  which is plotted as a function of \mh
in Figure~\ref{sm-cls}a. Values of $1-CL_b$ below $5.7\times 
10^{-7}$, indicated by the
horizontal full line, corresponding to a 5 standard deviation
fluctuation of the background, are considered to be in the discovery 
region. The dotted line shows the
expectation in the presence of a signal at a given \mh mass; its 
crossing with the $5\sigma$ line at
106.3~\GeVcc indicates the range of sensitivity of the presently 
available data to a discovery.

A 95\% confidence level lower limit on the Higgs mass may be set by
identifying the mass region where $CL_s < 0.05$, as shown in
Figure~\ref{sm-cls}. The \cls estimates the probability that the data
are compatible with  background and signal.  The median limit 
expected in the absence of a
signal is 109.1~\GeVcc\ and the limit observed by combining the LEP data
is 107.9~\GeVccf. The inclusion of systematic errors, together with their
correlations, has decreased the limits by approximately 100~MeV/$c^{2}$

\section{Conclusions and perspectives}

Searches for the Standard Model Higgs have been performed combining 
data collected by
the four LEP experiments at centre of mass energies between 192 and 
202~\GeVf, with a
an approximate integrated luminosity of 900~\pbinvf. These data have 
been combined with
data collected at lower energies. No statistically significant excess 
has been observed
when compared  to the Standard Model background predictions and a 
lower bound on the mass
of the Higgs boson has been set at 107.9~\GeVcc at 95~\% confidence
level.\\
This summer LEP will run at the highest possible energy. LEP is
expected to reach center of mass energies of around 206~\GeV.
 Prospects~\cite{Eilam} predict a combined sensitivity 
of around 114~\GeVcc for an integrated luminosity of 40~\pbinv per experiment 
at $\sqrts=206~\GeV$ and taking in account the data collected at  \sqrts=202~\GeVf. The discovery limit will be around 112~\GeVccf.
\section*{Acknowledgements}
I would like to thank the organisers of the XXXV$^{\mathrm{th}}$ Rencontres de Moriond
for the nice atmosphere of the Conference. I am grateful to the
Higgs searchers of the four LEP Collaborations and the LEP Higgs working group
for their help in preparing this talk. I would like to  thank
Achille Stocchi for his help and comments in the preparation of the talk 
and the careful reading of these proceedings.

\end{document}